\documentstyle[12pt,aasms4]{article}

\def\arcsecpoint{$''\!.$}
\def\deg{$^{\rm o}$}

\received{}
\accepted{}

\slugcomment{submitted to {\it The Astrophysical Journal}}

\lefthead{Kraemer, Crenshaw, Yaqoob, ...}
\righthead{Kinematics and Physical Conditions in Markarian 509 II.}

\begin{document}

\title{The Kinematics and Physical Conditions of the Ionized
Gas in Markarian 509. II. STIS Echelle Observations.\altaffilmark{1}}

\author{S. B. Kraemer\altaffilmark{2},
D. M. Crenshaw\altaffilmark{3},
T. Yaqoob\altaffilmark{4,6},
B. McKernan\altaffilmark{4},
J.R Gabel\altaffilmark{2},
I.M. George\altaffilmark{5,6},
T.J. Turner\altaffilmark{5,6},
\& J.P. Dunn\altaffilmark{3}}

\altaffiltext{1}{Based on observations made with the NASA/ESA Hubble Space
Telescope. STScI is operated by the Association of Universities for Research in
Astronomy, Inc. under the NASA contract NAS5-26555. }
  
\altaffiltext{2}{Catholic University of America,
NASA's Goddard Space Flight Center, Code 681,
Greenbelt, MD  20771; stiskraemer@yancey.gsfc.nasa.gov.} 

\altaffiltext{3}{Department of Physics and Astronomy,
Georgia State University, One Park Place South SE, 
Suite 700, Atlanta, GA 30303} 

\altaffiltext{4}{Department of Physics and Astronomy, Johns
Hopkins University, Baltimore, MD 21218.}

\altaffiltext{5}{Joint Center for Astrophysics, University of
Maryland, Baltimore County, 1000 Hilltop circle, Baltimore, MD
21250.}

\altaffiltext{6}{Laboratory for High Energy Astrophysics, 
NASA's Goddard Space Flight Center, Code 662,
Greenbelt, MD  20771.} 

\begin{abstract}

We present observations of the UV absorption lines in the luminous
Seyfert 1 galaxy 
Mrk 509, obtained with the 
medium resolution ($\lambda$/$\Delta\lambda$ $\approx$ 40,000) echelle gratings
of the Space Telescope Imaging Spectrograph on the
{\it Hubble Space Telescope}. The spectra reveal the presence
of eight kinematic components of absorption in Ly$\alpha$, C~IV, and N~V, at 
radial velocities of $-$422, $-$328, $-$259, $-$62, $-$22, $+$34,
$+$124, and $+$ 210 km s$^{-1}$ with respect to an emission-line redshift
of z $=$ 0.03440, seven of which were detected
in an earlier {\it Far Ultraviolet Spectrographic Explorer (FUSE)} spectrum. 
The component at $-$22 km s$^{-1}$ 
also shows absorption by Si~IV. The covering factor and velocity width
of the Si~IV lines were lower than those of the higher ionization lines for 
this component,
which is evidence for two separate absorbers at this velocity.   
We have calculated photoionization models to match the UV column densities
in each of these components. Using the predicted O~VI column densities,
we were able to match the O~VI profiles observed in the {\it FUSE}
spectrum. Based on our
results, none of the UV absorbers can produce the X-ray
absorption seen in simultaneous {\it Chandra} observations; therefore, there must be more highly ionized gas in the
radial velocity ranges covered by the UV kinematic components. 

\end{abstract}


\section{Introduction}

Since the launch of the {\it International Ultraviolet Explorer (IUE)}, it has been
known that the UV spectra of Seyfert 1 galaxies show absorption 
lines intrinsic to their nuclei 
(Ulrich 1988). With the advent of the {\it Hubble Space Telescope (HST)},
it is now understood that intrinsic absorption is a common phenomenon,
present in more than half of the well-studied Seyfert 1 galaxies (Crenshaw
et al. 1999). Among those Seyferts that show absorption, high ionization lines
such as N~V $\lambda\lambda$1238.8, 1242.8
and C~IV $\lambda\lambda$1548.2, 1550.8 are always present, along with
Ly$\alpha$, while lower ionization lines, such as 
Si~IV $\lambda\lambda$1393.8, 1402.8 and Mg~II
$\lambda\lambda$2796.3, 2803.5, are less common. Typically, the absorption lines are 
blueshifted (by up to 2100 km s$^{-1}$) with respect to the
systemic velocities of the host galaxies, indicating net radial outflow.
Although there are examples of strong UV absorption lines near systemic velocities,
some are found in highly inclined host galaxies and, therefore, are most likely formed 
in gas within the plane of the host galaxy (Crenshaw et al. 2001; Crenshaw et al. 2002).  
Among the blue-shifted absorbers, the ionic columns are highly variable, which may be the result
of changes in response to the ionizing continuum (cf. Krolik \&
Kriss 1997; Shields \& Hamann 1997) or transverse motion (Crenshaw \&
Kraemer 1999). Variability is suggestive of the proximity of
the absorbers to the central active nucleus, since it may result from the high
densities, hence short recombination timescales, or high transverse velocities, similar to those
inferred for the emission-line gas close to the active nucleus. Based on density constraints, it has been shown that 
in at least two sources, NGC 4151 and NGC 3516, much 
of the absorbing gas may lie within a fraction of a parsec from the central 
source (Kraemer et al. 2001a; Kraemer et al. 2002).
Another indication of small radial distances is the low line-of-sight covering factors derived from
some UV absorption lines (Kraemer et al. 2002; Gabel et al. 2002).
In addition to the UV absorbers, the presence of intrinsic absorption, typically in the form of 
bound-free edges of O~VII and O~VIII, has been detected in the X-ray 
spectra of a similar fraction of Seyfert 1 galaxies (Reynolds 1997; 
George et al. 1998). Most recently, spectra obtained with the {\it Chandra
X-ray Observatory (CXO)}
have revealed that X-ray absorption lines associated with this
material are also blue-shifted (Kaastra et al. 2000; Kaspi et al. 2000,
2001). Although it has been argued that some fraction of the UV absorption
arises in the same gas responsible for the X-ray absorption (Mathur, Elvis
\& Wilkes 1995, 1999; Crenshaw \& Kraemer 1999; Kriss et al.
2000; Kraemer et al. 2002), the connection between the X-ray and UV absorption is 
complex. In fact, there is often a wide range in ionization states in
gas which may be at the same radial velocities (Kaspi et al. 2002). Further
evidence for this is the co-spatial X-ray and optical line emission seen
in the narrow-line regions (NLR) of Mrk 3 (Sako et al. 2000), NGC 1068 (Ogle et al.
2002), and NGC 4151 (Ogle et al. 2000). 

Mrk 509 is a highly luminous (L$_{h\nu~>~13.6 eV}$ $\sim$ 10$^{45}$ erg s$^{-1}$;
see Kriss et al. [2000]; Yaqoob et al. [2002a]) Seyfert 1 galaxy.
Phillips et al. (1983) determined that the nucleus is surrounded by
two distinct components of extended ionized gas: a low ionization component, with
a velocities indicative of rotation in the galactic disk, and a high ionization
component, with velocities blue-shifted with respect to those in the disk.
Phillips et al. suggested that this high 
ionization component is part of
an outflowing shell of gas. If gas is distributed within a 
bicone centered on the active nucleus, as appears to be the case in other
Seyfert galaxies (Crenshaw \& Kraemer 2000; Crenshaw at al. 2000; Ruiz
et al. 2001), the axis of the outflow must be roughly parallel
to our line-of-sight (see Fig. 8 in Phillips et al.).  There is extended broad Balmer line emission 
(FWZI $>$ 15,000 km s$^{-1}$), presumably scattered light from the unresolved 
broad line region, in an axisymmetrical distribution
surrounding the active nucleus (Mediavilla et al. 1998), consistent
with the geometry suggested by Phillips et al. (1983). Based on recent X-ray
observations of Mrk 509 with {\it XMM-Newton}, Pounds et al. (2001) argued that
the inclination of the putative accretion disk within the 
active nucleus is $<$ 30\deg, which also
fits with the proposed geometry.

Two kinematic components of intrinsic Ly$\alpha$ and C~IV absorption, at
$\sim$ $-$420 and $+$40 km s$^{-1}$, were detected by York et al.
(1984) in high-dispersion {\it IUE} spectra of Mrk 509. In their
{\it HST}/Faint Object Spectrograph observations $\sim$ 12 yr later, Crenshaw, Boggess,
\& Wu (1995) found the same components, as well as N~V absorption at these
velocities. York et al. (1984) and Crenshaw et al. (1995) suggested that these
lines formed in extended regions of ionized gas in the host galaxy. 
Kriss et al. (2000) obtained a high resolution 
($\lambda$/$\Delta\lambda$ $\approx$ 15,000) {\it Far Ultraviolet Spectroscopic
Explorer (FUSE)} spectrum over the 
wavelength range 915 -- 1185 \AA\ on 1999 November 9, 11 that shows intrinsic 
absorption in the lines of O~VI $\lambda\lambda$1031.9, 1037.6, C~III 
$\lambda$977.0, and the H~I Lyman lines, consisting of seven, relatively narrow
(FWHM $\leq$ 63 km s$^{-1}$), 
kinematic components which are clustered into two groups, at $-$370 km s$^{-1}$ and near the systemic velocity. 
They suggested that the former was associated with the extended, blue-shifted ionized gas
observed by Phillips et al. (1983). In addition to the UV absorption, {\it ASCA}
spectra showed strong evidence for the presence of an X-ray warm absorber 
(Reynolds 1997; George et al. 1998). 

We have obtained simultaneous {\it CXO}/High Energy Transmission Grating (HETG) and
{\it HST}/Space Telescope Imaging Spectrograph (STIS) medium resolution spectra of Mrk 509 on 2001 April 13. 
We note that there are no previous {\it HST} 
high-resolution ($\lambda$/$\Delta\lambda$ $\geq$ 10,000) UV spectra, except 
for a Goddard High-Resolution Spectrograph (GHRS) observation of the intrinsic 
L$\alpha$ absorption (Crenshaw et al. 1999), which is highly saturated (see 
section 2.2). 
The analysis of the X-ray spectra are discussed in Yaqoob et al. (2002a)
and Yaqoob et al. (2002b; hereafter Paper I). In Paper I, we describe
the details of the ionizing continuum and the physical nature of the X-ray warm 
absorbing gas, including model predictions of ionic column densities.
In the present paper, we present our analysis of the STIS spectra and
the results of photoionization modeling of the UV absorbers.  
The paper is organized as follows: in Section 2 we describe the observations, 
and give the details of the measurement of the intrinsic lines, 
in Section 3 we detail the photoionization modeling of the 
absorbers, in Section 4 we discuss the implications of the
result, and Section 5 gives our summary. 

\section{Observations and Data Analysis}

\subsection{New STIS Observations}
We obtained STIS echelle spectra of the nucleus of Mrk 509
through a 0\arcsecpoint2 x 0\arcsecpoint2 aperture on 2001 April 13.
The details of these observations are listed in Table 1. 
We reduced the STIS spectra using the IDL software developed at NASA's 
Goddard Space Flight Center for the STIS Instrument Definition Team.
The data reduction includes a procedure to remove the background light from 
each order using a scattered light model devised by Lindler (1998). The 
individual orders in each echelle spectrum were spliced together in the 
regions of overlap.

Figure 1 shows portions of the STIS spectra where intrinsic absorption 
is detected in the UV. The spectra are normalized (as 
described below) and are plotted as a function of the radial velocity, 
relative to a systemic 
redshift of z $=$ 0.03440 from the optical emission lines (Crenshaw et al. 
1999), since H~I 21-cm measurements are not available. We have adopted the 
numbering of components given by Kriss et al. (2000). Each of their components 
is present in N~V $\lambda\lambda$ 1238.8, 1242.2 and C~IV $\lambda$ 1548.2, 
1550.8, and we identify a new component (4$'$) in the blue wing of component 
4; the new detection is likely due to the higher spectral resolution of STIS.
Although these components are completely saturated and blended together in 
Ly$\alpha$, they are all undoubtedly present, since they are present and
resolved in the 
{\it FUSE} spectra of Ly$\beta$.
We have also detected Si~IV $\lambda\lambda$ 
1393.8, 1402.8 in component 4, but not in any other component. We have not 
detected lower ionization lines (e.g., Si 
III $\lambda$1206.5, C~II $\lambda$1334.5, Mg II $\lambda\lambda$ 
2796.3, 2803.5) in any component in the STIS spectra.

Figure 1 shows that the absorption components are clustered in two 
groups in radial velocity; components 2 and 4 from the two clusters are the 
strongest in the high-ionization lines of 
C~IV and N~V. None of the components is 
highly saturated, since the long-wavelength member of each doublet is clearly 
not as deep as the short-wavelength member (which has twice the oscillator 
strength) in each case. The saturated Ly$\alpha$ absorption covers the same 
velocity range as these two clusters, and there is no evidence for H~I 
absorption outside of the two clusters. We note that our velocity scale is 
offset 
from that of Kriss et al. (2000) by 52 km s$^{-1}$; they use a systemic 
redshift of z $=$ 0.03457 (cz $=$ 10,365 km s$^{-1}$) from the nuclear [O~III] 
emission in Phillips et al. (1993), whereas we use a systemic redshift of z 
$=$ 0.03440 (cz $=$ 10,313 km s$^{-1}$) to be consistent with our earlier 
measurements (Crenshaw et al. 1995, 1999). We note that components 5 -- 7 are 
not necessarily redshifted with respect to the host galaxy, since there is a 
tendency for the [O III] $\lambda\lambda$ 4959, 5007 emission in Seyfert 1 
galaxies to be slightly blueshifted with respect to the systemic velocity 
determined from H~I 21-cm emission (Crenshaw et al. 1999).

\subsection{Measurements and Observational Results}

The procedures we used to measure the intrinsic absorption lines 
follow those of Crenshaw et al. (1999). To determine the shape of the 
underlying emission, we fit a cubic spline to regions on either side of the 
absorption, and then normalized the absorption profiles by dividing the 
observed spectra by the spline fits. 
We determined the covering factor C$_{los}$, which is the fraction 
of continuum plus emission that is occulted by the absorber in our line of 
sight, using the technique of Hamann et al. (1997) for doublets. 
Due to the modest signal-to-noise of the spectra and problems with blending 
of 
the components, we can only determine C$_{los}$ reliably in the cores of the 
strongest, relatively unblended lines. Thus, we used both C~IV and N~V 
doublets to determine C$_{los}$ for the cores of components 1, 2, 4, and 6.
To determine the column densities, we converted each normalized profile 
to optical depth as a function of radial velocity, and integrated across the 
profile, as described in Crenshaw et al. (1999). For the components with 
strong evidence for nonunity covering factors, we determined the true optical 
depths using the measured C$_{los}$ values and the formalism of Hamann et al. 
(1997). For the other components, we assumed C$_{los}$ $=$ 1.

Table 2 gives the measured radial velocity centroids, widths (FWHM), covering 
factors, and associated one-sigma errors for each kinematic component. 
These values represent the means and standard deviations 
determined from the measurements of the individual C~IV and N~V profiles. For 
component 4, the Si~IV profiles yield a very different width and covering 
factor compared to the C~IV and N~V profiles. We take this as an indication 
that there are two different physical components at this radial velocity, 
which we call ``4 (high)'' and ``4 (low)'', for high- and low-ionization 
components, respectively. 

The covering factors for components 4 (high) and 6 are consistent with a 
value of one, although we cannot rule out the possibility that they are 
slightly lower. Components 1, 2, and 4 (low) all show strong evidence for a
nonunity covering factor, whereas the covering factors for the weaker 
components 3, 4$'$, 5, and 7 cannot be reliably determined. Unocculted 
emission from the narrow-line region (NLR) in this small aperture cannot 
explain 
the nonunity covering factors, since this emission would be centered near 
zero 
velocity and would decrease away from line center. The simplest 
explanation for the partial covering is that the absorbers associated with 
components 1, 2, and 4 (low) 
are not occulting the entire continuum source and/or broad-line region (BLR) 
as seen in projection on the sky.

We list the ionic column densities for each component in Table 3.
Components 4 (high) and 4 (low) required special treatment. For component 4 
(low), our photoionization models indicate very large columns of C~IV and N~V 
(Section 3.2). To see if these columns were consistent with the 
observed profiles, we simulated the C~IV and N~V profiles by assuming the 
same profile and covering factor as those for Si~IV (C$_{los}$ $=$ 0.32). The 
simulated C~IV and N~V profiles are plotted in Figure 1, and demonstrate that 
the large columns for component 4 (low) in Table 6 can indeed by hidden in 
the 
observed profiles. However, column densities spanning a wide range of values 
could be hidden, due to the small covering factor of this component, and so 
we 
have no observationally determined values for component 4 (low) in Table 3.
To determine the column densities for component 4 (high), we assumed that the 
C~IV and N~V lines for component 4 (low) were indeed saturated, and 
therefore form pedestals from which the component 4 (high) optical depths can 
be measured against.

\subsection{The Far-UV Light Curve of Mrk 509}

To place the STIS observations in context with previous far-UV observations, 
we have retrieved all of the previous low-resolution spectra of Mrk 509 
obtained in the far-UV. Table 4 lists these
spectra (i.e., those not listed in Table 1), which were obtained by the 
{\it IUE}, the Hopkins Ultraviolet 
Telescope ({\it HUT}), and {\it HST}/FOS. 
We retrieved the most recently processed versions of these spectra from the 
Multimission Archive at STScI. We measured their continuum 
fluxes by averaging the points in a bin centered at 
1385~\AA\ (observed frame) with a width of 40 \AA. We determined the one-sigma 
flux errors from the standard deviations of the averages; for the {\it IUE} 
spectra, this technique is known to overestimate the errors (Clavel et al. 
1991), and we scaled the errors for these observations by a factor of 0.5 to 
ensure that observations taken on the same day agreed to within the errors (on 
average).

Figure 2 shows that the far-UV continuum of Mrk~509 varied 
dramatically over a span of 23 years. The continuum light curve is clearly 
undersampled, as evidenced by the large variations over $\sim$60 days near JD 
2,448,200. It can be seen that the STIS observation was obtained when the 
far-UV continuum was in a low state: F$_{\lambda}$(1385\AA) $=$ 5.3 ($\pm$0.4) 
x 10$^{-14}$ ergs s$^{-1}$ cm$^{-2}$ \AA$^{-1}$. Interestingly, 
the simultaneous {\it CXO} observations show that Mrk 509 was in an average
X-ray state at this time (Yaqoob et al. 2002a). Although the {\it FUSE} 
observation does not contain this wavelength region, we can compare it with 
our STIS observation by looking at the continuum in the region of overlap. At 
1180 \AA, the continuum fluxes are 6.0 ($\pm$0.7) and 6.2 ($\pm$1.0) x 
10$^{-14}$ ergs s$^{-1}$ cm$^{-2}$ \AA$^{-1}$ for the {\it FUSE} and STIS 
spectra, respectively, indicating that these spectra were taken in a similar 
low state. It would be interesting to obtain STIS
observations of Mrk 509 in a high state, given strong evidence for absorption 
variations in response to continuum changes in NGC 4151 (Kraemer et al. 2001a) 
and NGC 3516 (Kraemer et al. 2002).

\section{Photoionization Models}

\subsection{Inputs to the Photoionization Models}

 Photoionization models for this study were generated using the
 code CLOUDY90 
 \footnote{The X-ray models described in Paper I were generated with
 the code XSTAR; we found no significant differences in the ionic column densities
 predicted by the two codes.} (Ferland et al. 1998). We have modeled the absorbers
 as matter-bounded slabs of atomic gas, irradiated
 by the ionizing continuum radiation emitted by the central source.
 As per convention, the models
 are parameterized in terms of the ionization parameter,
 \begin{equation}
U = {1\over{4\pi~r^2~n_H~c}}~ \int^{\infty}_{\nu_0} ~\frac{L_\nu}{h\nu}~d\nu,
\end{equation}
where $L_{\nu}$ is the frequency dependent luminosity of the ionizing continuum,
$r$ is the distance between the cloud and the ionizing source and
$h\nu_{0}$ = 13.6 eV, and $n_{H}$ is the 
 number density of hydrogen atoms at the illuminated face of 
 the slab (note that, in general, $r$ and $n_{H}$
 cannot be determined independently in absorption line studies). The models are also
 parameterized in terms of the total
 hydrogen column density, N$_{\rm H}$ ($=$ N$_{\rm H~I}$ $+$ N$_{\rm H~II})$\footnote{We 
 use N$_{\rm XM}$ to denote the ionic column density, where ``X'' is the atomic 
 symbol and ``M'' is the ionization state.}. For the models, we have assumed only thermal 
 broadening, since the absorption lines are quite narrow (see Table 2) and
 widths greater than thermal (which are $\sim$ a few km s$^{-1}$) are likely to be due to the
 superposition of unresolved kinematic components (note that the resolution of the
 STIS E140M echelle grating is $\sim$ 7 km s$^{-1}$ at 1550 \AA).  
 The details of the ionizing continuum and element abundances assumed for these models 
 are given in Paper I. The absorbing gas is assumed to be free of cosmic dust.

\subsection{Model Results}

  The ``best-fit'' values of U and N$_{\rm H}$ (determined by matching the measured values of N$_{\rm N~V}$, N$_{\rm C~IV}$ and 
  N$_{\rm Si~IV}$)
  for the models are listed in Table 5, and
  the predicted ionic column densities are given in Table 6. Based on the
  models, none of the UV absorbers has a sufficient column density to significantly
  modify the intrinsic continuum at wavelengths above the Lyman limit. Hence, even though several of the
  components have covering fractions near unity, there will be no strong
  effects due to the screening of one component by another. 
  For components 1, 2, 3, 4$'$, 5, 6, and 7, the models match the observed N$_{\rm N~V}$ 
  and N$_{\rm C~IV}$ to well within the measurement errors. 
  Our predictions for N$_{\rm C~III}$ for these components are in good agreement 
  with Kriss et al. (2000), with the expcetion of a factor of
  $\sim$ 8 underprediction for component 1 and a factor of $\sim$ 2 overprediction
  for component 3. The latter is likely to the different line widths assumed for
  this component, while the discrepancy for component 1 may be evidence for a 
  second low ionization component at this velocity (also, see Fig. 2 in Kriss 
  et al. [2000]). The models also underpredict N$_{\rm H~I}$ for components
  1, 3, 5, 6, and 7. However, the additional H~I columns may arise in
  the X-ray absorber (see Section 3.3).
  
  For the low ionization fraction of component 4, the model predictions match the 
  measured value for N$_{\rm Si~IV}$ and the upper limits for N$_{\rm Si~III}$ and N$_{\rm C~II}$. 
  As noted in Section 2.2, the model predictions for N$_{\rm C~IV}$ and N$_{\rm N~V}$ 
  for the low ionization component were then 
  used to determine the respective columns from the high ionization
  gas associated with component 4, which we were then able to model 
  successfully. The large H~I column predicted by the low ionization component
  is in rough agreement with that derived from  Ly$\beta$ by 
  Kriss et al. (2000), which suggests that the low ionization gas covers the
  UV continuum source but not the BLR clouds (which contribute a negligible
  amount of Ly$\beta$ emission; see discussion in Gabel et al. [2002]).
  Although the low ionization model predicts N$_{\rm C~III}$ $\sim$ 30 
  times greater than that derived by Kriss et al. (2000), the apparent
  discrepancy is clearly due to large covering factor assumed by those
  authors (C$_{los}$ $=$ 0.92), which is appropriate for the high ionization
  component but not the gas in which the C~III line forms.

Our model predictions for the O~VI column densities are much larger than those 
measured by Kriss et al. (2000) for components 1, 3, 4, by factors of 3 
-- 12, and lower in component 5 by a factor of 4. Of course, this could be the result of variability in these components, 
but we consider that unlikely, given that the {\it FUSE} and {\it HST}/STIS 
observations were obtained at a similar continuum state. Inspection of the 
{\it FUSE} O~VI profiles shows that they are highly saturated, and the 
components could potentially harbor much larger columns than those
measured by Kriss et al. To test the 
compatability of our model predictions with the {\it FUSE} observation, we 
have generated a simulated O~VI $\lambda$1031.9 profile in the following 
manner. First, we isolated the optical depth profile of each kinematic 
component in N~V$\lambda$1238.8. Then, we scaled each profile so that they 
would yield the predicted O~VI columns from our models. Finally, we combined 
the optical depths and converted them to normalized flux (as a function of 
radial velocity) using the covering factors of each component given by Kriss 
et al. There are two reasons for using the Kriss et al. covering factors, 
rather than the ones we have determined. First, it is possible for two 
absorption lines to have different covering factors, even if they come from 
the same gas, due to different contributions to the underlying emission from 
the continuum source and BLR (for a demonstration of this effect, see Gabel et 
al. [2002]). Second, the {\it FUSE} aperture (30$''$ x 30$''$) is much larger 
than the STIS aperture that we used (0\arcsecpoint2 x 0\arcsecpoint2), and it 
is possible that more unocculted emission (e.g., from the NLR, a scattering 
region, or hot stars in the host galaxy) is admitted by the large aperture, 
which could contribute to the difference in the derived covering factors
at low radial velocities, particularly for
components 4 (high ionization) and 6.

Our simulated O~VI profile is shown in Figure 3. Comparison with Figure 2 in 
Kriss et al. (2000) shows that it provides a very good match to the observed 
O~VI $\lambda$1031.9 profile. There is a small discrepancy in that the flux 
between components 5 and 6 is a little high in the simulated profile, but that 
could be due to a slight underestimate of the N~V optical depth in the wings of 
these components. We conclude that a direct comparison of model and observed 
O~VI column densities is not feasible, due to heavy saturation of the 
absorption components, but that the O~VI column densities from our models are 
at least consistent with the observed O~VI profile.

\subsection{The Nature of the UV Absorbers}

Based on our model predictions, the components of UV absorption in Mrk 509
have smaller column densities and are, on average, in a lower state of ionization
than the strongest components in the Seyfert 1 galaxies NGC 5548 
(log U $=$ $-$0.62, N$_{\rm H}$
$=$ 2.0 x 10$^{20}$ cm$^{-2}$; Crenshaw et al. 2002, in
preparation),
NGC 3783 (log U $=$ $-$0.19, N$_{\rm H}$ $=$ 1.5 x 10$^{21}$ cm$^{-2}$; Kraemer et al. 2001b), and 
NGC 3516 (log U $=$ $-$0.72, N$_{\rm H}$ $=$ 1.4 x 10$^{21}$ cm$^{-2}$; Kraemer et al. 2002). 
Interestingly, they are similar in column density and FWHM to the weak, narrow components 
detected in NGC 4151, which we had determined were at least tens of 
pcs from the central source (Kraemer et al. 2001a). The combination of high 
and low ionization
absorbers within a single kinematic component (component 4) has been
seen in NGC 3783 (Kraemer et al. 2001b; Gabel et al. 2002) and NGC 3516
(Kraemer et al. 2002). If these components are at the same radial
distance, there must be density inhomogeneities within the absorbing gas,
with the lower ionization absorption occuring in denser gas.
The fact that the low ionization component has a lower covering factor  
than the high ionization component is also evidence for different densities,
since the denser component is likely more compact.

Since, summed over the eight model kinematic components, the 
total N$_{\rm OVII}$ is $=$ 3.2 x 10$^{16}$ cm$^{-2}$, 
or about 10\% of the total observed O~VII column, none of the UV absorbers can be identified
with the X-ray absorbing gas described in Paper I. This is contrary to the suggestion
by Kriss et al. (2000) that, based on the large O~VI column, component 5 
must strongly contribute to the X-ray absorption. However, as discussed
in Section 3.2, the O~VI column predicted by our models is
consistent with the {\it FUSE} spectrum. Also, a model with physical parameters
determined by Kriss et al. (2000) for this component (log U $=$ $-$0.43, N$_{\rm H}$
$=$ 5.0 x 10$^{20}$ cm$^{-2}$) predicts a N~V column of
$\sim$ 10$^{15}$ cm$^{-2}$, which is clearly inconsistent with the
STIS spectrum. The discrepancy may result from the difficulty in deriving accurate lower
limits to column densities from such heavily saturated lines.
 
Models of the X-ray absorber (Paper I)  predict detectable C IV and N V
column densities of 1.2 x 10$^{13}$ cm$^{-2}$ and 8.5 x 10$^{13}$ cm$^{-2}$,
respectively.  However, as discussed above, our models of the UV components,
which assume a single physical component for each (except for component 4),
do not predict substantial X-ray absorption. Thus, the UV columns associated
with the X-ray absorber are likely to be ``hidden'' in one or more of the UV
components. If the X-ray absorber is coincident with only one UV component,
the C IV column could  be hidden in any kinematic component, but the N~V
column could only be hidden in components 1, 2, 4$'$, or 4. Alternatively,
the C~IV and N~V columns from the X-ray absorber could be spread out amongst
some or all of the kinematic components. Thus, it is clear that the X-ray
absorption occurs within the velocity ranges covered by the two clusters of
UV absorption, since otherwise we would have uniquely detected its UV
columns. The velocity coverage of the X-ray absorption in the {\it CXO}
spectrum is consistent with this conclusion (Paper I).
The Ly$\alpha$ profile in Figure 1 is also consistent with the conclusion
that the X-ray absorption must fall in the velocity ranges covered by the UV
absorbers, since we expect a substantial H~I column from the X-ray absorber
(1.4 x 10$^{15}$ cm$^{-2}$, see Paper I), and yet we see no H~I absorption in
the Ly$\alpha$ profile (Figure 1) that is outside of the two clusters of UV
absorption.
Based on the {\it FUSE} spectra, Kriss et al. (2000) derived a total
N$_{\rm H~I}$ $\approx$ 4.7 x 10$^{15}$ cm$^{-2}$ (excluding component 4,
in which the Ly$\beta$ line may form principally in the low ionization component).
The UV models predict a total N$_{\rm H~I}$ $=$ 1.6 x 10$^{15}$ cm$^{-2}$
(excluding components 4 and 4$'$), which leaves 3.1 x 10$^{15}$ cm$^{-2}$
unaccounted for. This is roughly consistent with the X-ray model prediction,
and further evidence for additional gas at the same velocities of
a number of the UV components. It is interesting that there is also strong
evidence for a wide range in ionization state at the same velocities in the
intrinsic UV and X-ray absorbers in the Seyfert 1 galaxy NGC 3783 (Kaspi et al.
2002).

\section{Discussion}

In general, intrinsic UV (Crenshaw et al. 1999) and X-ray (Kaspi et al. 2000, 2001;
Kaastra et al. 2000) absorption lines detected in Seyfert 1 galaxies are
blue-shifted, indicative of radial outflow. However, in the case of 
Mrk 509, the clustering of 
component velocities near systemic and the presence of red-shifted gas (although
this is somewhat uncertain; see Section 2.1) differ from the more obvious
cases of radial outflow discussed by Crenshaw et al. (1999). 
Kriss et al. (2000) suggested that this may be the result of the absorbers at
radial distances of $\sim$ 300 pc, co-located with the low ionization
gas detected by Phillips et al. (1983) and moving 
primarily transversely to our line-of-sight. Indeed,
it has been suggested that UV absorption lines form
in disk-driven winds (e.g., Proga, Stone, \& Kallman 2000). If this is the
case, the low velocities we observe in Mrk 509 require a viewing angle
roughly orthogonal to the direction of the flow, similar to the geometry
suggested by Kriss et al. (2000). This scenario is consistent with the
evidence that the accretion disk in Mrk 509 is viewed roughly face-on 
(Pounds et al. 2001). If the widths of the
UV absorption lines detected in Seyfert 1s are the result of superposition
of multiple kinematic components, viewing an outflow stream from this
vantage point might help explain the narrowness of the lines seen in Mrk 509. 
Nevertheless, there are some problems with this hypothesis. Since Mrk 509 is
a high luminosity Seyfert 1, disk-wind models predict that the outflow
will be more confined to the disk than in lower luminosity AGN (Konigl \&
Kartje 1994; Murray et al. 1995). The axisymmetric distribution of the reflected broad Balmer
line emission in Mrk 509 can be interpreted as evidence for an obscuring
torus or wind that covers only a small solid angle (Mediavilla et al. 1998),
which is consistent with a ``flattened'' wind. However, most of the UV 
components have covering factors near unity, which may be difficult to achieve
with a flattened wind viewed against a roughly face-on disk, depending
on the location of the continuum source relative to the base of the wind.
Specifically, the ``launch-pad'' of the wind must be {\it interior} to
the UV continuum source.
 
Alternatively, it is possible that the observed velocities of the UV absorbers
are close to their true radial velocities. In this case, the
absorbers may not be part of a disk-driven flow and are, instead, gas that has
been elevated off the disk and, therefore, may be more likely to occult the
continuum and BLR when the disk is viewed face-on. Although disk-wind
models include such a component, both the X-ray and UV absorbers detected
in Mrk 509 are too neutral to be the ``hitch-hiker'' gas proposed by
Murray et al. (1995) to shield the disk-driven winds or the highly-ionized gas in the inner part of the
wind modeled by Proga et al. (2000), since, in both cases, these components may have ionization
parameters $\geq$ 10. 
Krolik \& Kriss (1995) proposed that 
UV absorbers are high density knots embedded in a thermally expanding,
highly ionized wind, which would provide
a natural explanation for the kinematic association of the UV and
X-ray absorbers. However, this model predicts radial velocities 
similar to the disk-driven winds ($>$ several hundred km s$^{-1}$),
hence would still require a special geometry to explain 
the low radial velocities and large values of C$_{los}$.
A simpler scenario would have the absorbers at large
radial distances, e.g. 100's of pcs as suggested by Kriss
et al. (2000). As noted in Section 3.3, similar
UV absorbers were found at large distances in NGC 4151
(Kraemer et al. 2001a). Furthermore, studies of the narrow-line region
kinematics indicate a strong drop in radial velocities in emission-line
gas at distances $>$ few hundred pcs (Crenshaw \& Kraemer 2000; Crenshaw
et al. 2000; Ruiz et al. 2001), so it is not necessarily surprising that
the absorbers within the NLR would have similar kinematics.
However, better determination of the radial distances of these absorbers
requires constraints on the gas density, which can only be achieved
via variability studies.

\section{Summary}

We have used UV medium resolution echelle spectra obtained with {\it HST}/STIS
and X-ray spectra obtained with {\it CXO}/HETG to study the physical conditions 
in the intrinsic absorbers in the Seyfert 1 galaxy Mrk 509. The analysis
of the X-ray warm absorber is given in Paper I, while here we describe the
nature of the UV absorbers. We have determined the following:

1. We have detected 8 separate kinematic components in Ly$\alpha$, C~IV, and
N~V, seven of which were previously detected in a {\it FUSE} spectrum. 
Component 4 also shows the presence of Si~IV. We derived a C$_{los}$  $\approx$
0.32 from the Si~IV lines, as opposed to $\approx$ 0.98 from the high
ionization lines, which we interpret as evidence for two separate 
physical components
at this velocity, similar to that observed in NGC 3783 (Kraemer et al. 2001b)
and NGC 3516 (Kraemer et al. 2002). 

2. Contrary to the analysis in Kriss et al. (2000), none of the UV absorbers possesses 
a sufficiently high column density or ionization parameter to be identified with the 
X-ray warm absorber. However, based on the warm absorber models presented in Paper I, 
it is likely that highly ionized gas is kinematically associated with one or more of the UV components.

3. There is evidence suggesting that accretion disk in Mrk 509 is 
viewed roughly face-on. Although disk-wind models would predict
low velocities for the UV absorbers with this geometry, it is clear
that they cannot account for the large covering factors derived for
several of the components. Alternatively, the properties of the
absorbers may be consistent with outflow perpendicular to the surface of a
disk. Furthermore, it is plausible that the absorbers lie within the NLR 
of Mrk 509, at significant radial distances from the central source, and hence
a part of a different population of intrinsic absorbers than that detected
in several other Seyfert 1 galaxies. The UV continuum is currently in a low-state, while
the X-ray contimuum flux is near its historical average (Yaqoob et al. 2002a);
therefore spectra taken as the source brightens (over a timeframe of several
yrs) will provide a means to constrain the densities and, thus, the 
radial distances of the absorbers.

\acknowledgments

Support for proposal 8877 was provided by NASA through a grant from the Space 
Telescope Science Institute, which is operated by the Association of 
Universities for Research in Astronomy, Inc., under NASA contract NAS 5-26555. 
Some of the data presented in this paper were obtained from the Multimission 
Archive at the Space Telescope Science Institute (MAST). Support for MAST for 
non-HST data is provided by the NASA Office of Space Science via grant 
NAG5-7584 and by other grants and contracts. We would like to thank
Jose Ruiz for useful discussions. We thank an anonymous referee for their 
comments.

\clearpage

\figcaption[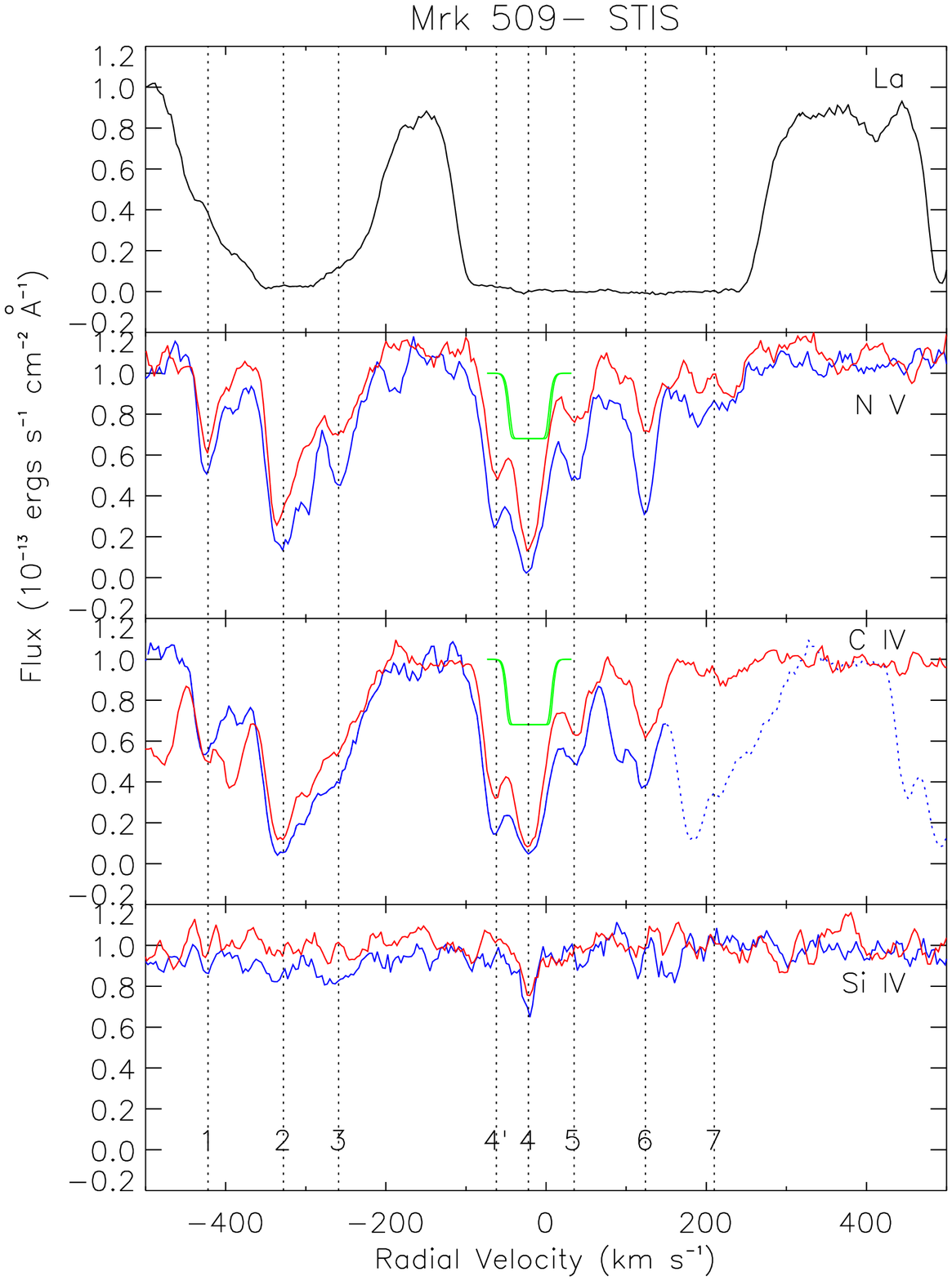]{ Portions of the STIS echelle spectra of Mrk~509, showing 
the intrinsic absorption lines in different ions. Normalized fluxes are 
plotted as a function of radial velocity, relative to an emission-line 
redshift of z $=$ 0.03440. 
For the doublets, the short wavelength member is plotted in blue, and the long 
wavelength member is plotted in red.
The kinematic components are identified with vertical 
dotted lines at their measured positions.
Areas where a component from one doublet member are heavily contaminated by 
another component from the other member are plotted as dotted lines.
Simulated N~V and C~IV profiles for component 4 (low), as described in the 
text, are plotted in green.}

\figcaption[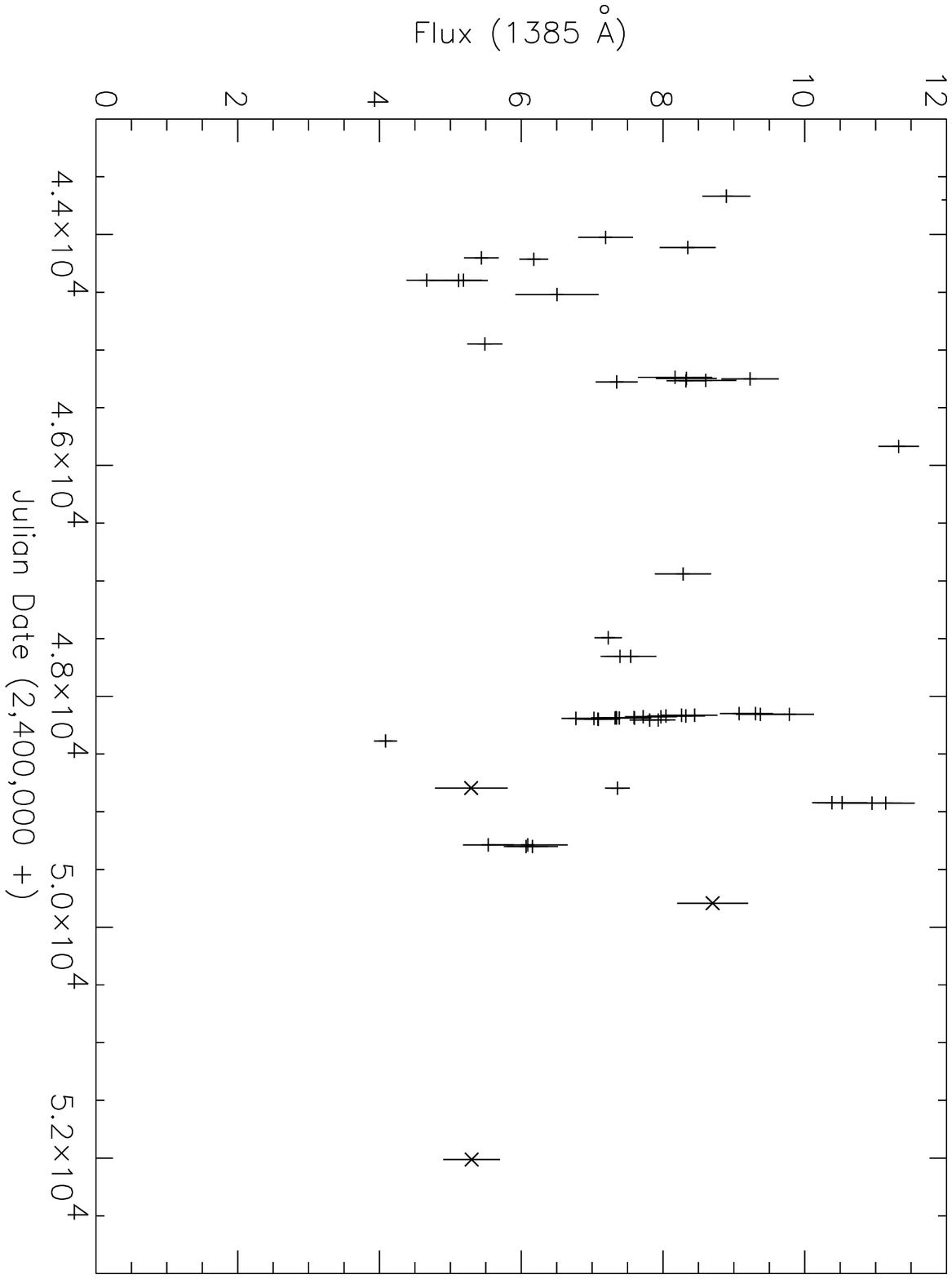]{Far-UV continuum light curve of Mrk~509. Fluxes (in
10$^{-14}$ ergs 
s$^{-1}$ cm$^{-2}$ \AA$^{-1}$) at 1385~\AA\  are plotted as a function of 
Julian date. The pluses are from {\it IUE}, and the X's are from the other 
satellites; vertical lines indicate the error bars ($\pm$ one $\sigma$).}

\figcaption[f2.eps]{Simulated O~VI $\lambda$1031.9 profile, based on the 
optical depth profiles from N~V, the covering factors of Kriss et al. (2000), 
and the predicted O~VI columns from the model of each component, as described 
in the text. The normalized flux is plotted as a function of radial velocity, 
relative to an emission-line redshift of z $=$ 0.03440.}

\newpage
\begin{deluxetable}{ccccll}
\tablecolumns{6}
\footnotesize
\tablecaption{{\it HST}/STIS High-Resolution Spectra of Mrk 509}
\tablewidth{0pt}
\tablehead{
\colhead{Instrument} & \colhead{Grating} & \colhead{Coverage} &
\colhead{Resolution} & \colhead{Exposure} & \colhead{Date} \\
\colhead{} & \colhead{} & \colhead{(\AA)} &
\colhead{($\lambda$/$\Delta\lambda$)} & \colhead{(sec)} & \colhead{(UT)}
}
\startdata
STIS &E140M &1150 -- 1730 &46,000 &7600   &2001 April 13 \\
STIS &E230M &2275 -- 3120 &30,000 &2700   &2001 April 13 \\

\enddata
\end{deluxetable}

\begin{deluxetable}{lrcr}
\tablecolumns{4}
\footnotesize
\tablecaption{Absorption Components in Mrk~509}
\tablewidth{0pt}
\tablehead{
\colhead{Comp.} & \colhead{Velocity$^{a}$} & \colhead{FWHM} & 
\colhead{C$_{los}$$^{b}$}\\
\colhead{} &\colhead{(km s$^{-1}$)} &\colhead{(km s$^{-1}$)} & \colhead{}
}
\startdata
1        &$-$422 ($\pm$2)  & 28 ($\pm$6) & 0.52 ($\pm$0.05) \\
2        &$-$328 ($\pm$2)  & 49 ($\pm$8) & 0.91 ($\pm$0.06) \\
3        &$-$259 ($\pm$2)  & 41 ($\pm$2) & --------------- \\
4$'$     &$-$62  ($\pm$2)  & 32 ($\pm$3) & --------------- \\
4 (high) &$-$22 ($\pm$1)   & 52 ($\pm$5) & 0.98 ($\pm$0.06) \\
4 (low)  &$-$21 ($\pm$2)$^{c}$   & 21 ($\pm$3)$^{c}$ & 0.32 ($\pm$0.03)$^{c}$ 
\\
5        &$+$34 ($\pm$2)   & 35 ($\pm$3) & --------------- \\
6        &$+$124 ($\pm$3)  & 29 ($\pm$6) & 1.02 ($\pm$0.04) \\
7        &$+$210 ($\pm$11) & 53 ($\pm$9) & --------------- \\
\tablenotetext{a}{Velocity centroid for a systemic redshift of z $=$ 0.03440.}
\tablenotetext{b}{Covering factor in the line of sight.}
\tablenotetext{c}{Determined from Si~IV, rather than N~V and C~IV.}
\enddata
\end{deluxetable}

\begin{deluxetable}{lccccccc}
\tablecolumns{8}
\footnotesize
\tablecaption{Measured Ionic Column Densities (10$^{14}$ cm$^{-2}$)}
\tablewidth{0pt}
\tablehead{
\colhead{Comp.} & \colhead{N~V} & \colhead{C~IV} & 
\colhead{Si~IV} & \colhead{Si~III} & \colhead{C~II}
}
\startdata
1    & 1.16 $\pm$0.19 &  0.70 $\pm$ 0.10 & $<$0.03 & $<$0.01 & $<$0.05 \\
2    & 2.40 $\pm$0.19 &  2.78 $\pm$ 0.29 & $<$0.03 & $<$0.01 & $<$0.05 \\
3    & 0.58 $\pm$0.10 &  0.92 $\pm$ 0.33 & $<$0.03 & $<$0.01 & $<$0.05 \\
4$'$ & 0.84 $\pm$0.17 &  0.81 $\pm$ 0.08 & $<$0.03 & $<$0.01 & $<$0.05 \\
4 (high) & 1.67 $\pm$ 0.20 &  1.32 $\pm$ 0.32 & $<$0.03 & $<$0.01 & $<$0.05 \\
4  (low) & ------------  &  ------------  & 0.21 $\pm$ 0.05 & $<$0.05 & 
$<$0.20 \\
5    & 0.43 $\pm$0.10 &  0.29 $\pm$ 0.07 & $<$0.03 & $<$0.01 & $<$0.05 \\
6    & 0.63 $\pm$0.10 &  0.39 $\pm$ 0.05 & $<$0.03 & $<$0.01 & $<$0.05 \\
7    & 0.36 $\pm$0.09 &  0.10 $\pm$ 0.03 & $<$0.03 & $<$0.01 & $<$0.05 \\

\enddata
\end{deluxetable}

\begin{deluxetable}{lcrl}
\tablecolumns{4}
\footnotesize
\tablecaption{Low-Resolution FUV Spectra of Mrk 509}
\tablewidth{0pt}
\tablehead{
\colhead{Instrument/} & \colhead{Coverage} &
\colhead{Resolution} & \colhead{Date} \\
\colhead{Grating} & \colhead{(\AA)} &
\colhead{($\lambda$/$\Delta\lambda$)} & \colhead{(UT)}
}
\startdata
IUE SWP    &1150 -- 1978 &$\sim$250 &1978 June 7 -- 1993 November 9$^{a}$\\
FOS G130H & 1150 -- 1605 &$\sim$1200 &1992 June 21 \\
HUT  &~820 -- 1840 &$\sim$450 &1995 March 16 \\

\tablenotetext{a}{See the IUE Merged Log at http://archive.stsci.edu/iue.}
\enddata
\end{deluxetable}

\begin{deluxetable}{lccc}
\tablecolumns{4}
\footnotesize
\tablecaption{Photoionization Model Parameters}
\tablewidth{0pt}
\tablehead{
\colhead{Comp.} & \colhead{N$_{\rm H}$ (cm$^{-2}$)} & \colhead{log(U)} 
& \colhead{log($\xi$)$^{a}$}}
\startdata
1 & 1.03 x 10$^{19}$ & -0.82 & 0.67\\
2 & 8.30 x 10$^{18}$ & -1.31 & 0.18 \\
3 & 1.77 x 10$^{18}$ & -1.48 & 0.01\\
4$'$ & 3.40 x 10$^{18}$ & -1.19 & 0.30\\
4 (high) & 9.35 x 10$^{18}$ & -1.02 & 0.47 \\
4 (low) & 3.40 x 10$^{19}$ & -1.70 & -0.21 \\
5 & 3.86 x 10$^{18}$ & -0.82 & 0.67 \\
6 & 6.20 x 10$^{18}$ & -0.78 & 0.71 \\
7 & 4.00 x 10$^{19}$ & -0.16 & 1.33 \\
\tablenotetext{a}{
For comparison to XSTAR models (see Paper I), we also list the ionization
parameter
$\xi$ = ${1\over{r^2~n_H}}$~ $\int^{13.6 keV}_{\nu_0}$ ~${L_\nu}$~d$\nu$.
Based on the model SED, $\xi$ $=$ (0.0321)U.}
\enddata
\end{deluxetable}

\begin{deluxetable}{lcccccccc}
\tablecolumns{9}
\footnotesize
\tablecaption{Predicted Ionic Column Densities$^{a}$ (10$^{14}$ cm$^{-2}$)}
\tablewidth{0pt}
\tablehead{
\colhead{Comp.} & \colhead{H~I} & \colhead{O~VI} & \colhead{N~V} 
& \colhead{C~IV} & \colhead{Si~IV} & \colhead{C~III} & \colhead{Si~III} 
& \colhead{C~II}
}
\startdata
1    &  2.58 & 22.9 & 1.16 & 0.75 & -- & 0.02 & -- & --  \\
 & (5.70) & (1.90) & (1.16) & (0.70) & ($<$0.03) & (0.16) & ($<$0.01) &
($<$0.05)\\
2 & 7.25 & 20.2 & 2.40 & 2.77 & -- & 0.26 & -- & -- \\    
 & (5.30) & (13.0) & (2.40) & (2.78) & ($<$0.03) & (0.24) & ($<$0.01) & 
 ($<$0.05) \\
3 & 2.39 & 3.35 & 0.59 & 0.91 & -- & 0.13 & -- & --\\ 
   & (9.30) & (1.20) & (0.58) & (0.92) & ($<$0.03) & (0.05) & ($<$0.01) & ($<$0.05) \\
4$'$ & 2.22 & 9.02 & 0.84 & 0.82 & -- & 0.06 & -- & -- \\ 
 & & & (0.82) & (0.81) & ($<$0.03) &  & ($<$0.01) & ($<$0.05) \\
4 (high) & 3.87 & 24.8 & 1.67 & 1.33 & -- & 0.06 & -- & -- \\
   & -- & (12.0) & (1.67) & (1.32)  & -- & -- & -- & -- \\
4 (low) & 78.1 & 36.9 & 11.0 & 28.3 & 0.21 & 6.58 & 0.05 & 0.07 \\
        & (60.0) & --  &  -- & --  & (0.21) & (0.21) & ($<$0.05) & 
($<$0.20) \\
5 & 0.97 & 8.60 & 0.43 & 0.28 & -- & 0.01 & -- & --\\  
 & (2.70) & (32.0) & (0.43) & (0.29) & ($<$0.03) & ($<$0.02) & ($<$0.01) & ($<$0.05) \\
6  & 1.40 & 13.0 & 0.63 & 0.39 & -- & 0.01 & -- & -- \\  
  &  (12.0) & (12.0) & (0.63) & (0.39) & ($<$0.03) & ($<$0.02) & ($<$0.01) & 
  ($<$0.05) \\
7 & 1.43 & 15.7 & 0.37 & 0.10 & -- & -- & -- & -- \\  
  & (12.0) & (9.40) & (0.36) & (0.10) & ($<$0.03) & ($<$0.02) & ($<$0.01) & ($<$0.05) \\
\tablenotetext{a}{Measured values are listed in parentheses on the second line.
The values for O~VI, C~III, and H~I are from Kriss et al. (2000).}
\enddata
\end{deluxetable}


\clearpage
\vskip3.0in
\begin{figure}
\plotone{f1.eps}
\\Fig.~1.
\end{figure}

\clearpage
\vskip3.0in
\begin{figure}
\plotone{f2.eps}
\\Fig.~2.
\end{figure}

\clearpage
\vskip3.0in
\begin{figure}
\plotone{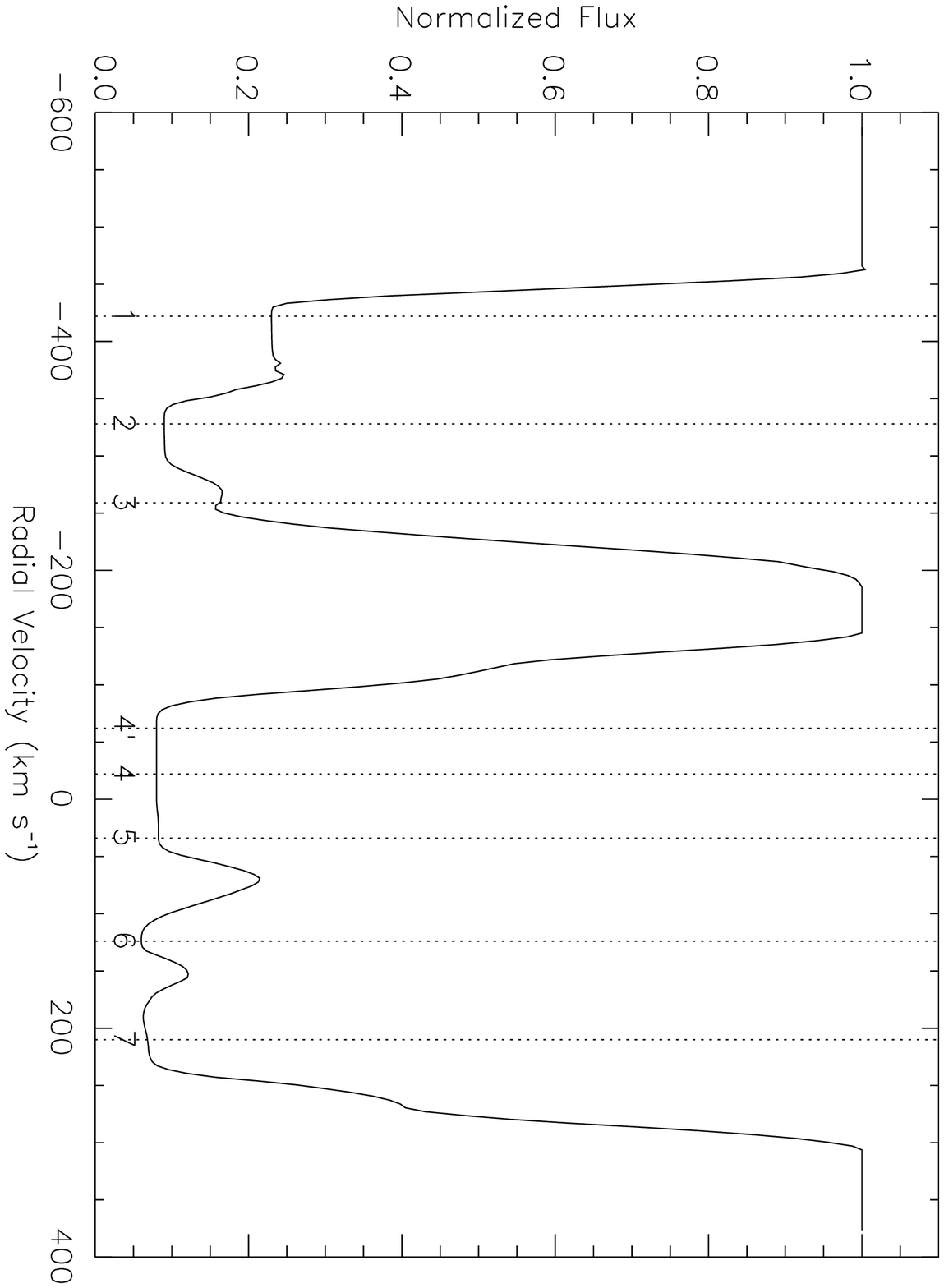}
\\Fig.~3.
\end{figure}

\acknowledgments

\end{document}